\documentclass[aps,pre,twocolumn,groupedaddress]{revtex4}
\usepackage{graphicx}

\begin{document}
\title{Using reservoir computers to distinguish chaotic signals}
\author{T. L. Carroll}
\email{Thomas.Carroll@nrl.navy.mil}
\affiliation{US Naval Research Lab, Washington, DC 20375}

\date{\today}

\begin{abstract}
Several recent papers have shown that reservoir computers are useful for analyzing and predicting dynamical systems. Reservoir computers have also been shown to be useful for various classification problems. In this work, a reservoir computer is used to identify one out of the 19 different Sprott systems. An advantage of reservoir computers for this problem is that no embedding is necessary. Some guidance on choosing the reservoir computer parameters is given. The dependance on number of points, number of reservoir nodes and noise in identifying the Sprott systems is explored.

\end{abstract}
\pacs{ 05.45.-a, 05.45.Tp}

\maketitle

\section{Introduction}
Describing chaotic signals is difficult because of their complex nature. If an experiment produces a chaotic signal, some way to describe the signal is necessary to detect changes in the experiment. There have been a number of methods published for comparing or identifying chaotic signals \cite{farmer1987, sugihara1990, casdagli1989, hunt2007, brown1994, carroll2016, carroll2017a, gilmore1998}, but most of these methods require embedding the signal in a phase space, which requires knowledge of the embedding dimension and delay. Phase space embeddings are also sensitive to noise, as individual points are displaced by noise in multiple dimensions, so inter-point distances are not accurate. There are other methods for characterizing attractors, such as fractal dimension, Lyapunov exponents, linking numbers, etc \cite{abarbanel1993}. These methods are commonly used because in theory they are invariant under orientation preserving diffeomorphisms, so that a change in the embedded variable or the embedding method should not change the measurement. In practice, there are well known problems when applying these standard methods to real data; see, for example, \cite{bradley2015}. 

There has been quite a bit of recent work on using reservoir computers to model and predict chaotic systems \cite{lu2018,zimmerman2018,antonik2018,lu2017,van_der_sande2017} so it is known that reservoir computers are useful for analyzing chaotic signals. A reservoir computer is simply a high dimensional dynamical system that is driven by a signal to be analyzed. Usually the dynamical system is created by connecting a set of nonlinear nodes in a network so that the entire dynamical system has a stable fixed point. The dynamical system then responds to the input signal of interest, acting as a nonlinear filter. Training of the reservoir computer comes about by forming a linear combination of many signals from the dynamical system to fit a training signal; for example, in \cite{lu2017}, the dynamical system is driven by the Lorenz $x$ signal, and a set of signals from the dynamical system is fit to the Lorenz $z$ signal. The fit coefficients are saved. In computational mode, the same dynamical system is then driven by a Lorenz $x$ signal with different initial conditions. Using the previously fit coefficients to make a linear combination of signals from the dynamical system, the $z$ signal corresponding to the particular $x$ signal is reproduced.

The reservoir computer does introduce additional complexity in that the dynamical system typically contains from 100 to 1000 nodes. Long term research focuses on implementing reservoir computers as analog systems \cite{van_der_sande2017, lukosevicius2009}, creating an advantage in terms of computational speed. Another complication in applying reservoir computers is that there is no theory of how reservoir computers operate, so choosing parameters for a reservoir computer to solve a particular problem proceeds by trial and error.

In this paper, I begin by describing a reservoir computer and how to train the computer. I then choose particular parameters for the reservoir computer using signals from the Sprott B chaotic system \cite{sprott1994}. Next I describe how to create a coefficient vector that is characteristic of a particular Sprott system, and I show how to use these cofficient vectors to determine from which Sprott system a particular signal originated, and I characterize the error performance of this signal identification.

\section{Reservoir Computing}
Reservoir computing is a branch of machine learning  \cite{jaeger2004,lukosevicius2009,manjunath2013}. A reservoir computer consists of a set of nonlinear nodes connected in a network. The set of nodes is driven by an input signal, and the response of each node is recorded as a time series. A linear combination of the node response signals is then used to fit a training signal. Unlike other types of neural networks, the network connecting the nonlinear nodes does not vary; only the coefficients used to fit the training signal vary.

The reservoir computer used in this work is described by
\begin{equation}
\label{res_comp}
\frac{{d{\bf{R}}}}{{dt}} = \lambda \left[ {\alpha {\bf{R}} + \beta {{\bf{R}}^2} + \gamma {{\bf{R}}^3} + {\bf{AR}} + {\bf{W}}s\left( t \right)} \right].
\end{equation}
$\bf{R}$ is vector of node variables, ${\bf A}$ is a sparse matrix indicating how the nodes are connected to each other, and ${\bf W}$ is a vector that described how the input signal $s(t)$ is coupled to each node. The constant $\lambda$ is a time constant, and there are $M$ nodes. For all the simulations described here, $\alpha, \beta$ and $\gamma$ are set to make the network stable, that is the network has a stable fixed point with a large basin of attraction.

The particular reservoir computer used here is arbitrary, and other types of nodes can also be used. The main requirements for a reservoir computer are that the nodes are nonlinear and that the network of nodes has a stable fixed point, so that in the absence of an input signal the network does not oscillate \cite{manjunath2013}.

\begin{figure}
\centering
\includegraphics[scale=0.25]{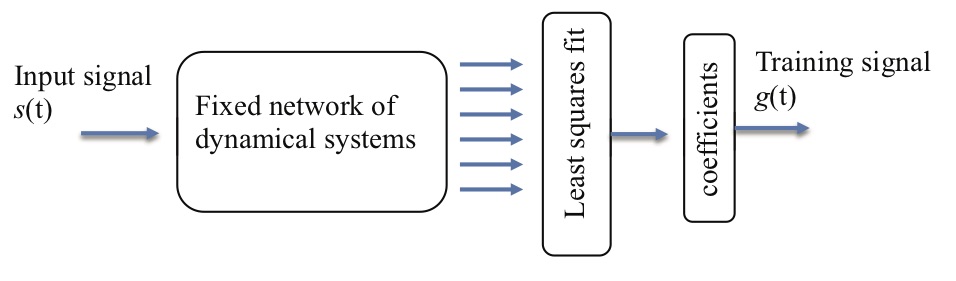} 
  \caption{ \label{reservoir_computer}
Block diagram of a reservoir computer. The input signal $s(t)$ drives a fixed network of dynamical nodes. The time varying signal from the nodes are fit to the training signal $g(t)$ by a least squares fit.}.
  \end{figure}

Figure \ref{reservoir_computer} is a block diagram of a reservoir computer. 
  
To train the reservoir computer, an input signal $s(t)$ and a training signal $g(t)$ were chosen and equation \ref{res_comp} was numerically integrated. The first part of the response of the reservoir computer was discarded as a transient, and the next $N$ time series points $r_i(t), i=1 ... M$ from each node were combined in a $N \times (M+1)$ matrix
\begin{equation}
\label{fit_mat}
\Xi   = \left[ {\begin{array}{*{20}{c}}
{{r_1}\left( 1 \right)}&{{r_1}\left( 2 \right)}& \ldots &{{r_1}\left( N \right)}\\
 \vdots &{}&{}&{}\\
{{r_M}\left( 1 \right)}&{{r_M}\left( 2 \right)}& \ldots &{{r_M}\left( N \right)}\\
1&1& \ldots &1
\end{array}} \right]
\end{equation}
The last row of $\Xi $ was set to 1 to account for any constant offset in the fit. The training signal is fit by
\begin{equation}
\label{train_fit_0}
g\left( t \right) = \sum\limits_{j = 1}^M {{c_j}{r_j}\left( t \right)} 
\end{equation}

or
\begin{equation}
\label{train_fit}
{g(t)} = \Xi  {\bf{C}}
\end{equation}
where $g(t) = \left[ {g\left( 1 \right),g\left( 2 \right) \ldots g\left( N \right)} \right]$ is the training signal.

The matrix $\Xi $ is decomposed by a singular value decomposition
\begin{equation}
\label{svd}
\Xi   = {\bf{US}}{{\bf{V}}^T}.
\end{equation}
where ${\bf U}$ is $N \times (M+1)$, ${\bf S}$ is  $N \times (M+1)$ with non-negative real numbers on the diagonal and zeros elsewhere, and ${\bf V}$ is $(M+1) \times (M+1)$.

The pseudo-inverse of $\Xi $ is constructed as
\begin{equation}
\label{pinv}
{\Xi  _{inv}} = {\bf{V}}{{\bf{S}}^{\bf{'}}}{\bf{U}}
\end{equation}
where ${\bf{S}}^{\bf{'}}$ is an $(M+1) \times (M+1)$ diagonal matrix, where the diagonal element $S^{'}_{i,i}=S_{i,i}/(S_{i,i}^2+\delta^2)$, where $\delta=1 \times 10^{-5}$ is a small number used for ridge regression to prevent overfitting. 

The fit coefficient vector is then found by
\begin{equation}
\label{fit_coeff}
{\bf{C}} = {\Xi  _{inv}}{g(t)}
\end{equation}.
  
The coefficient vector ${\bf C}$ will be used as a feature vector to identify individual signals. The difference between signals  $i$ and $j$ is computed as
\begin{equation}
\label{sprott_diff}
{\Delta _{ij}} = \sum\limits_{k = 1}^{M + 1} {\sqrt {{\bf{C}}_i^2\left( k \right) - {\bf{C}}_j^2\left( k \right)} } 
\end{equation}
where ${{\bf C}_i}^2(k)$ is the $k'th$ component of the coefficient vector for signal $i$.

The training error $E_T$ may be computed from
\begin{equation}
\label{train_err}
{E_T} = \frac{{\left\| {\Xi  {\bf{C}} - {g(t)}} \right\|}}{{\left\| {g(t)} \right\|}}.
\end{equation}
The training error is used as a measure of how well the training signal $\bf{G}$ may be reconstructed from the input signal ${s(t)} = \left[ {s\left( 1 \right),s\left( 2 \right), \ldots s\left( N \right)} \right]$. 

\section{Sprott systems}
Sprott \cite{sprott1994} found a family of 19 different chaotic systems defined by 3-dimensional ODE's with 1 or 2 quadratic nonlinearities. This group of signals is a useful test set for our signal comparison methods. 
 
 Each set of ODE's for the Sprott systemss was integrated using a 4th order Runge-Kutta integrator with a time step of 0.01. The integrator output was decimated by keeping every 50'th point to produce a time series. 
  
As an example, the Sprott B system was described by the differential equations
  \begin{equation}
  \label{sprott_b_eqn}
\begin{array}{l}
\frac{{dx}}{{dt}} = yz\\
\frac{{dy}}{{dt}} = x - y\\
\frac{{dz}}{{dt}} = 1 - xy
\end{array}
  \end{equation}.
  
\begin{figure}
 \centering
    \includegraphics[scale=0.4] {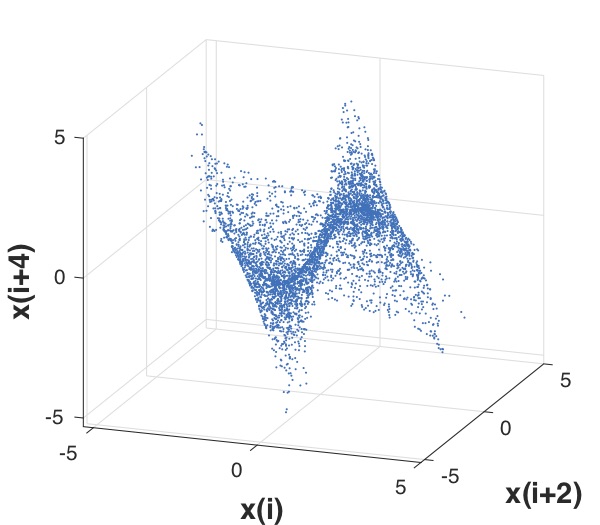} 
  \caption{ \label{sprottb}
Embedded time series signal for the Sprott B attractor with an embedding delay of 2.}
  \end{figure}

 Figure \ref{sprottb} is a plot of the embedded attractor for the Sprott B system.

\begin{figure}
 \centering
    \includegraphics [scale=0.4]{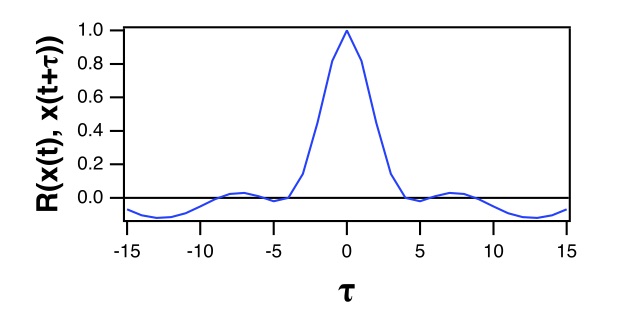} 
\caption{\label{sprottb_corr} Autocorrelation $R(x(t), x(t+\tau)$ for the $x(t)$ signal from the Sprott B system.}
\end{figure}

Figure \ref{sprottb_corr} is the autocorrelation of the $x(t)$ signal from the Sprott B system. The autocorrelation will be used in setting the parameters for the reservoir computer.

\section{Reservoir computer parameters}
\label{choose_par}
There is currently no theory for designing a reservoir computer to solve a particular problem, so parameter choice for eq. (\ref{res_comp}) must proceed by trial and error. The training error $E_T$ from eq. \ref{train_err} was used as a metric to judge the accuracy of the reservoir computer- the smaller $E_T$, the better the computer. The parameters that produce the smallest $E_T$ may not be the best parameters for calculating the difference $\Delta_{ij}$ between two signals, but optimizing for $\Delta_{ij}$ requires that we know in advance that the two signals are different. We may not know in advance if the signals are the same or different. 

First, the parameters $\alpha=-3$, $\beta=1$ and $\gamma=-1$ in eq. (\ref{res_comp}) were chosen so that the network was stable. The number of nodes was set at $M=100$. Next, the specific network matrix ${\bf A}$ and input coupling vector ${\bf W}$ were determined. The parameters ${\bf A}$ and ${\bf W}$ were determined by choosing 100 randomly selected ${\bf A}$ and ${\bf W}$ pairs and keeping the pair that yielded the lowest training error $T_E$.

For the determination of ${\bf A}$ and ${\bf W}$, the input signal $s(t)$ and the training signal $g(t)$ were set to the $x(t)$ variable from the Sprott B system. Both $s(t)$ and $g(t)$ were normalized to have a mean of 0 and a standard deviation of 1. The time constant $\lambda$ was set to an arbitrary value of 1.

The input signal $s(t)$ was 6000 points long, and after driving the reservoir, the first 1000 points from all signals were discarded as a transient. One hundred random realizations of ${\bf A}$ were generated from a uniform random distribution between $\pm 1$. The matrix ${\bf A}$ was sparse, with 20\% of its elements nonzero, and all nodes had at least one connection to other nodes. ${\bf A}$ was normalized so that the largest absolute value of the real part of its eigenvalues was 0.5. Another 100 random realizations of ${\bf W}$ were generated from a uniform random distribution between $\pm 0.5$.

The training error (eq. \ref{train_err}) was recorded for each random network configuration and the ${\bf A}$ and ${\bf W}$ pair that gave the lowest training error $E_T$ was retained as part of the optimum parameter set.

Next the value for the time constant $\lambda$, which determined the frequency response of the reservoir,  was set. The reservoir computer responds to a finite band of frequencies, so to make sure this band of frequencies was optimal for analyzing the Sprott system signals, the time constant was varied between 0.5 and 10 and the training error was computed with the input and training signals $s(t)$ and $g(t)$ equal to the  $x(t)$ signals for each of the 19 Sprott systems. 

Figure \ref{train_err} is a plot of the mean of the training error $E_T$ for all 19 different Sprott systems as the time constant $\lambda$ is varied.
\begin{figure}
 \centering
    \includegraphics [scale=0.4]{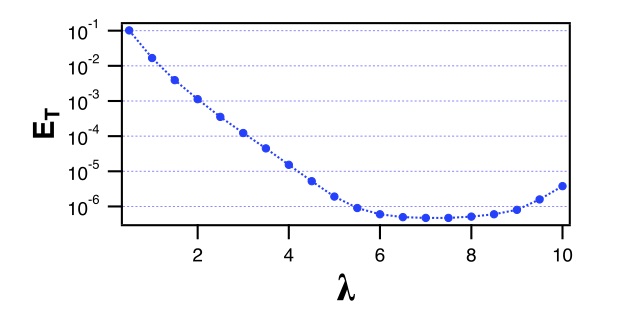} 
\caption{\label{train_err}Mean of the training error $E_T$ for all of the 19 different Sprott systems as the time constant $\lambda$ defined in eq. (\ref{res_comp}) is varied. The input signal $s(t)=x(t)$ for all 19 Sprott systems, and the training signal $g(t)$ was also equal to $x(t)$.}
\end{figure}

Figure \ref{train_err} shows that the mean of the training error for all the Sprott systems is small for $\lambda=7$, so the time constant for the reservoir computer in eq. (\ref{res_comp}) is set to 7. The minima for each of the individual Sprott systems occurred  roughly the same value of $\lambda$, so the minimum of the mean was a good approximation. If the value of $\lambda$ at which the minimum training error occurred was very different for different Sprott systems, then the signals could probably be distinguished by their frequency content alone, and no reservoir computer would be necessary.

\section{Signal identification}
It has been shown that reservoir computers are useful for signal identification or classification, for speech signals for example \cite{larger2012} or image recognition \cite{jalavand2018}. For signal fitting, the smallest training error will undoubtedly come when the training signal $g(t)$ is equal to the input signal $s(t)$; such is not the case for the error in identifying signals.

To identify the Sprott signals, the reservoir computer of eq. (\ref{res_comp}) was driven with the signal $x(t)$ from each of the Sprott systems, while the training signal $g(t)$ was set equal to $x(t+\tau)$. The reservoir computer of eq. (\ref{res_comp}) was numerically integrated with a 4th order Runge-Kutta integration routine with a time step of 0.1. The first 1000 time steps were discarded and the next 5000 time steps from each node were used to find the fitting coefficients ${\bf C}$ as in eq. (\ref{fit_coeff}). 

For each Sprott system, a time series consisting of 600,000 points of the $x(t)$ signal was generated and divided into 100 sections of 6000 points each. The reservoir computer node variables ${\bf R}(t)$ were all initialized to an initial value of 0. For each section, the reservoir computer was driven by the input signal $s(t)=x(t)$ and the first 1000 points of the network response variables $r_i(t), i=1 ... M$ were dropped to eliminate the transient. The fitting coefficients for each Sprott system for each of the 100 sections were found according to eqs. (\ref{fit_mat}-\ref{fit_coeff}).

For each of the 19 Sprott systems there were therefore 100 sets of coefficients ${\bf C_i(k)}$, where $i$ = A, B ... S indicates the particular Sprott system and $k=1, 2 ... 100$ indicates the section of the Sprott signal. 

The difference between two Sprott systems was defined in eq. (\ref{sprott_diff}) as $\Delta_{ij}$. The difference between two sections from two Sprott systems is
\begin{equation}
\label{sect_diff}
{\Delta _{ij}}\left( {{l_1},{l_2}} \right) = \sum\limits_{k = 1}^{M + 1} {\sqrt {{\bf{C}}_i^2\left( {{l_1},k} \right) - {\bf{C}}_j^2\left( {{l_2},k} \right)} } 
\end{equation}
where $l_1=1,2 ... 100$ and $l_2=1,2 ... 100$ indicate the different sections of the Sprott signals, $i$ and $j$ indicate the different Sprott systems, and $k=1,2 ... M+1$ indicates the particular component of the coefficient vector.

When comparing the Sprott systems, when sections $l_1$ and $l_2$ are compared, if the minimum value if $\Delta_{ij}(l_1,l_2)$ is not $\Delta_{ii}(l_1,l_2)$
\begin{equation}
\label{diff_max}
\min \left[ {{\Delta _{ij}}\left( {{l_1},{l_2}} \right)} \right] < {\Delta _{ii}}\left( {{l_1},{l_2}} \right)\;j \ne i,\;{l_1} \ne {l_2}
\end{equation}
then an error is recorded. The comparisons are made for all the coefficient vectors of all the sections of all 19 Sprott systems, and the probability $P_E$ of making an error in correctly identifying each Sprott system was recorded.

\begin{figure}
 \centering
    \includegraphics[scale=0.4] {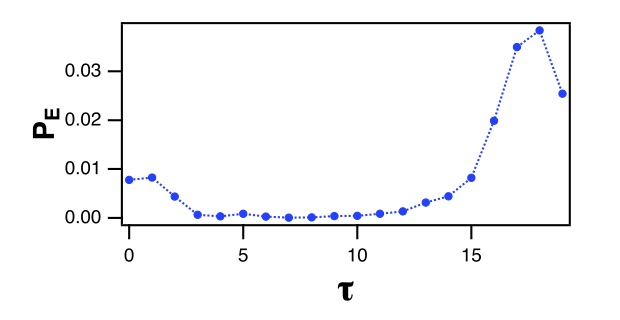} 
\caption{\label{var_delay} Probability of making an error $P_E$ in identifying the 19 Sprott systemss when the input signal to the reservoir computer $s(t)=x(t)$ and the training signal is $g(t)=x(t+\tau)$.}
\end{figure}

Figure \ref{var_delay} is a plot of the probability $P_E$ of making an error in identifying the Sprott systems as the delay $\tau$ in the training signal $g(t)=x(t+\tau)$ varies. Figure \ref{var_delay} shows that the smallest error in identifying the Sprott systems occurs when $\tau > 0$, so the training signal $g(t)$ does not match the input signal $x(t)$. The minimum error occurs for values of $\tau$ between 4 and 7.

The identification error probability is lower when the training signal $g(t)$ is delayed from the input signal $x(t)$ because the relation between the delayed signal and the non-delayed signal contains information unique to the particular chaotic system. Knowing what comes later in time for a particular signal gives more information than just knowing the signal at a particular time. To quantify this extra knowledge, the mutual information between the input signal $x(t)$ and the training signal $g(t)$ was computed. 

To compute the mutual information, each signal was transformed into a symbolic time series using the ordinal pattern method \cite{bandt2002}. Each signal was divided into windows of 4 points, and the points within the window were sorted to establish their order; for example, if the points within a window were 0.1, 0.3, -0.1 0.2, the ordering would be 2,4,1,3. Each possible ordering of points in $x(t)$ represented a symbol $\sigma_q(0), q=1 ... N_{s0}$, where $N_{s0}$ was the number of possible symbols. Each symbol in the delayed signal $x(t+\tau)$ was $\sigma_q(\tau), q=1 ... N_{s \tau}$. The probabilities $p(\sigma_q(0))$ and $p(\sigma_q(\tau))$ were found for each  symbol. The mutual information between the signal $x(t)$ and the delayed version $x(t+\tau)$ was
\begin{equation}
\label{minfo}
I\left( {0,\tau } \right) = \sum\limits_{q1 = 1}^{{N_{s0}}} {\sum\limits_{q2 = 1}^{{N_{s\tau }}} {p\left[ {{\sigma _{q1}}\left( 0 \right),{\sigma _{q2}}\left( \tau  \right)} \right]\log \left( {\frac{{p\left[ {{\sigma _{q1}}\left( 0 \right),{\sigma _{q1}}\left( \tau  \right)} \right]}}{{p\left[ {{\sigma _{q1}}\left( 0 \right)} \right]p\left[ {{\sigma _{q2}}\left( \tau  \right)} \right]}}} \right)} } .
\end{equation}

\begin{figure}
 \centering
    \includegraphics [scale=0.4]{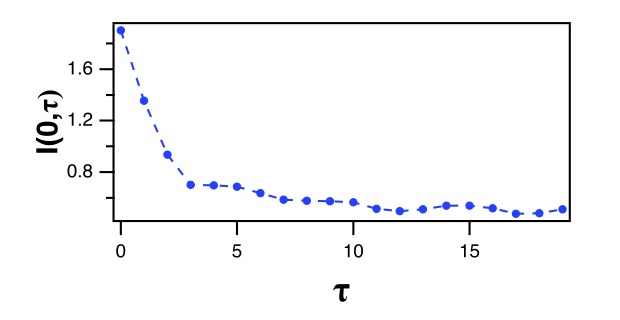} 
\caption{\label{mutual_info} Mean of the mutual information $I(0,\tau)$ between $x(t)$ and $x(t+\tau)$ for all 19 Sprott systems.}
\end{figure}

Figure \ref{mutual_info} is a plot of the mean of the mutual information between the input signal $x(t)$ and the delayed version $x(t+\tau)$ for all 19 Sprott systems. The mutual information between $x(t)$ and $x(t+\tau)$ decreases sharply for $\tau \le 3$ and then starts to level off. The delayed signal $x(t+\tau)$ has new information not present in $x(t)$, although the amount of new information does not increase as rapidly for $\tau > 3$.

The probability of identification error plot in fig. \ref{var_delay} increases for $\tau > 7$. The Sprott systems produce chaotic signals, so for long delays, $x(t+\tau)$ will be uncorrelated with $x(t)$. As an example, figure \ref{sprottb_corr} is a plot of the autocorrelation $R(x(t), x(t+\tau)$ for the $x(t)$ signal from the Sprott B system. The autocorrelation for the Sprott B system pictured in fig. \ref{sprottb_corr} first drops below 0 for $\tau=4$. For the other Sprott systems, the autocorrelation for the $x(t)$ signal drops below 0 for delays ranging from $\tau=2$ to $\tau=6$, except for the Sprott C system, where the autocorrelation doesn't drop below 0 until $\tau=33$, but the autocorrelation for the Sprott C system does have its first minimum at $\tau=5$. If the delay $\tau$ for the training signal $g(t)=x(t+\tau)$ is increased by too much, the training signal becomes uncorrelated with the input signal and the probability of identification error will increase.

\begin{figure}
 \centering
    \includegraphics [scale=0.4]{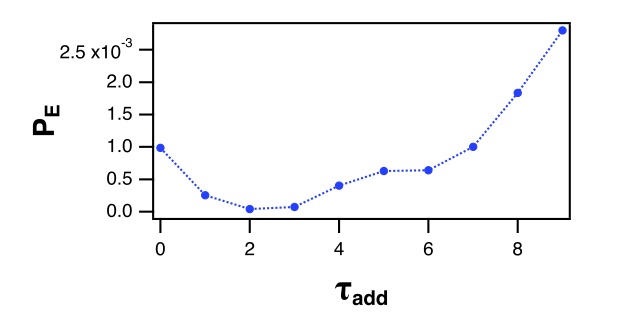} 
\caption{\label{tau_add} Probability of identification error $P_E$ as a function of the added delay $\tau_{add}$ for the 19 Sprott systems. The input signal was $s_i(t)=x_i(t)$, while the training signal was $g_i(t)=x_i(t+\tau_i+\tau_{add})$, where the index $i$ indicated the particular Sprott system. The delay $\tau_i$ was the delay for which the autocorrelation of the $x(t)$ signal from Sprott system $i$ first dropped below 0 or reached its first minimum.}
\end{figure}

To find the ideal delay for the training signal $g(t)=x(t+\tau)$ for each of the 19 Sprott systems, $\tau_i$, $i$=A,B ... S was set equal to the delay for which the autocorrelation for that signal first dropped below 0 or reached its first minimum. An additional delay $\tau_{add}$ was then added to each $\tau_i$, so each of the 19 training signals was $g_i(t)=x_i(t+\tau_i+\tau_{add})$. Figure \ref{tau_add} shows the probability of identification error $P_E$ as a function of the added delay for the Sprott systems. The identification error is minimized for $\tau_{add}=2$. It is evident that the optimum delay for the training signal $g(t)=x(t+\tau)$ is slightly greater than the delay for which the first minimum occurs in the autocorrelation function. Choosing this delay maximizes the new information provided by the training signal but keeps the training signal from becoming too uncorrelated with the input signal. This simple rule is similar to the conventional wisdom for choosing the delay window in a delay embedding, that the window length should be equal to the delay at which the first zero (or first minimum) in the autocorrelation is seen.

\subsection{Number of data points}
\begin{figure}
 \centering
    \includegraphics[scale=0.4] {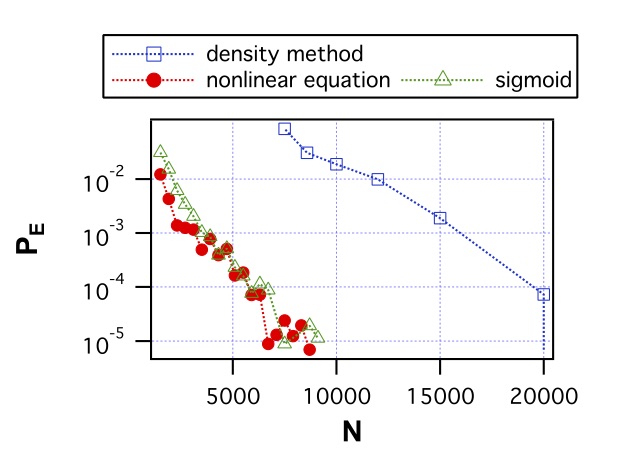} 
\caption{\label{err_npts} Probability of error $P_E$ as a function of number of points $N$ in the input time series $s(t)=x(t)$ for identifying the Sprott systems using a reservoir computer as described in eq. \ref{res_comp}, labeled as "nonlinear equation".  The input signal was $s(t)=x(t)$ for each of the Sprott systems, while the training signal $g(t)=x(t+\tau)$, where $\tau$ was the delay for the first 0 or first minimum of the autocorrelation function for each Sprott system plus 2 time steps. The figure also shows the probability of error for identifying the Sprott systems from the density method of \cite{carroll2016, carroll2017a}, labeled as the "density method" and from a reservoir computer using a sigmoid node described in eq. \ref{sig_map} ("sigmoid"). The number of points $N$ for the reservoir computers includes the 1000 point transient.}
\end{figure}

Figure \ref{err_npts} shows the error in identifying the 19 Sprott systems $P_E$ as a function of the total number of points used $N$. The total number of points includes the 1000 point transient. As a comparison, probability of error from the density method of \cite{carroll2016, carroll2017a} is also plotted. The reservoir computer method required fewer points to identify the Sprott systems, and it did not require that the signal be embedded in a phase space.

The error performance of the reservoir computer of eq. (\ref{res_comp}) was also compared to the error performance of a reservoir computer with a different type of node. The most commonly used node type in the reservoir computer literature is a sigmoid nonlinearity, so a reservoir computer described by eq. \ref{sig_map} was also simulated.

\begin{equation}
\label{sig_map}
{{\bf{R}}_i}\left( {n + 1} \right) = \alpha \left[ {\frac{1}{{1 + {e^{ - {{\bf{R}}_i}\left( n \right)}}}} + {\bf{AR}}\left( n \right) + {\bf{W}}s\left( t \right)} \right]
\end{equation}
The parameter $\alpha=0.35$. The network connection matrix ${\bf A}$ and the input coupling vector ${\bf C}$ were the same as in eq. (\ref{res_comp}), as were the input and training signals. The probability of error for identifying the 19 Sprott systems using the reservoir computer with a sigmoid nonlinearity is also plotted in fig. \ref{err_npts}. The probability of error when the sigmoid nonlinearity is used is approximately the same as when the nonlinear equation of eq. (\ref{res_comp}) is used.

\subsection{Number of nodes}
\begin{figure}
 \centering
    \includegraphics [scale=0.4]{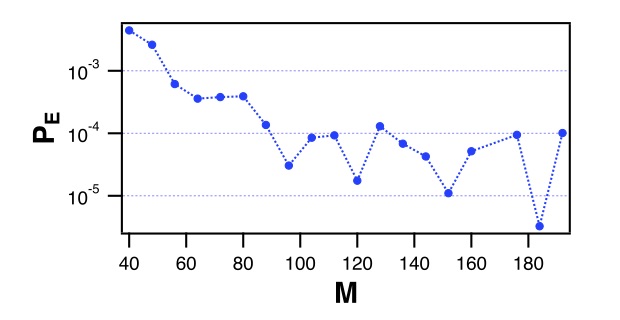} 
\caption{\label{err_vs_nodes} Probability of error $P_E$ in identifying the 19 Sprott systems using the reservoir computer of eq. (\ref{res_comp}) when the number of nodes $M$ in the network was varied.  The delay $\tau$ in the training signal $g(t)=x(t+\tau)$ was set to $\tau=\tau_i+2$, where $\tau_i$ the delay for which the autocorrelation for Sprott system $i$ first drops below 0 or has its first minimum.}
\end{figure}

It seems that the number of nodes $M$  in the reservoir computer should make a difference in the probability of error in identifying the Sprott systems, and figure \ref{err_vs_nodes} confirms this suspicion. Each time the number of nodes $M$ was changed, optimal values for the connection matrix ${\bf A}$ and the input vector ${\bf W}$ were determined as in section \ref{choose_par}.  The identification error $P_E$ as a function of the number of reservoir computer nodes $M$ is plotted in figure \ref{err_vs_nodes}.

Figure \ref{err_vs_nodes} shows that the scatter in the probability of identification $P_E$ is very large as the number of nodes increases. Most likely this scatter is caused by the fact that the network connection matrix ${\bf A}$ is not optimal. As the number of nodes $M$ increases, the number of elements in ${\bf A}$ increases as $M^2$, so the probability of randomly generating an optimum matrix  ${\bf A}$ from a fixed number of random realizations decreases, meaning that  ${\bf A}$ is less likely to be optimum as $M$ increases.

\subsection{Added noise}

\begin{figure}
 \centering
    \includegraphics[scale=0.4] {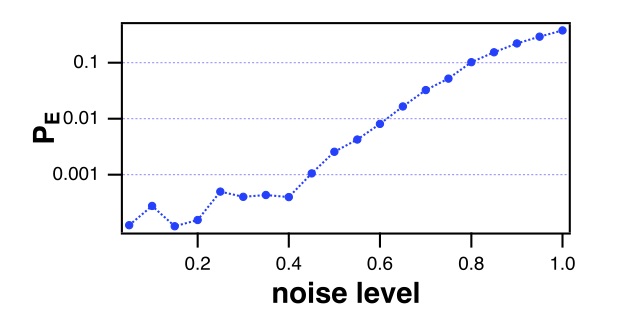} 
\caption{\label{err_vs_noise} Probability of error in identifying the 19 Sprott systems with added Gaussian white noise. The noise level is the ratio of the noise standard deviation to the signal standard deviation. The delay $\tau$ in the training signal $g(t)=x(t+\tau)$ was set to $\tau=\tau_i+2$, where $\tau_i$ the delay for which the autocorrelation for Sprott system $i$ first drops below 0 or has its first minimum. }
\end{figure}

Figure \ref{err_vs_noise} shows the probability of identification error $P_E$ for identifying the 19 Sprott systems as noise is added to the input signal; $s(t)=x(t)+\eta(t)$, where $\eta(t)$ is Gaussian white noise. The reservoir computer was described by eq. (\ref{res_comp}). The noise level in fig. \ref{err_vs_noise} is the ratio of the standard deviation of the noise signal to the standard deviation of $x(t)$. For each of the Sprott systems the delay $\tau$ used to determine the training signal $g(t)=s(t+\tau)$ was the delay for which the autocorrelation for that Sprott system first dropped below 0 (or had its first minimum) plus 2 time steps. The noise level on the training signal $g(t)$ is the same as the noise level on the input signal $s(t)$.

Figure \ref{err_vs_noise} shows that the reservoir computer of eq. (\ref{res_comp}) is robust to moderate amounts of added noise.

\section{Summary}
Reservoir computers are useful for identifying chaotic signals, as the example in this paper shows. Using a reservoir computer, it was possible to correctly identify signals from the 19 different Sprott systems with a error probability lower than that in a method that used density to identify chaotic systems \cite{carroll2016, carroll2017a}. An advantage of the reservoir computer method is that no embedding is required, so it isn't necessary to estimate dimension or delay.

One drawback to using reservoir computers is that there is no theory to guide the selection of reservoir computer parameters. The strategy used in this paper was to use choose input and training signals $x(t)$ and $g(t)$ and vary the reservoir parameters to minimize the training error $T_E$ (eq. \ref{train_err}). This approach gives reasonable parameters, but it is not optimum, since the parameters where the training error is minimized may not be the parameters that minimize the error in identifying different systems, $P_E$. 

The reservoir computer method does require numerically integrating a network of $M$ nonlinear systems, and as figure \ref{err_vs_nodes} shows, larger values of $M$ give lower error probabilities. Implementing a reservoir computer on a digital computer is slow, although the different nodes may be integrated in parallel. The real promise of reservoir computing is that the nodes may be implemented with analog systems, in which case speed increases over digital computing are possible.

\section{References}
\bibliography{using_reservoir_computers_v2}{}

\end{document}